\documentclass[reprint,aps,nofootinbib,superscriptaddress]{revtex4-2}

\usepackage{graphicx,amsmath,amsfonts,amssymb,slashed,float}
\usepackage{array,multirow}
\usepackage[utf8]{inputenc}
\usepackage[usenames,dvipsnames]{xcolor} 
\usepackage{dcolumn}
\usepackage{bbold,wasysym}
\usepackage{bm}
\usepackage[normalem]{ulem}
\usepackage[breaklinks=true,colorlinks=true,linkcolor=blue,urlcolor=blue,citecolor=blue,hyperfootnotes=false]{hyperref}
\usepackage{endnotes}
\usepackage{booktabs}
\usepackage{url}

\DeclareUnicodeCharacter{3B5}{$\epsilon$}
\DeclareUnicodeCharacter{2212}{\ensuremath{-}}
\DeclareUnicodeCharacter{039B}{\ensuremath{\Lambda}}

\newcommand{\Euclid}{\emph{Euclid}}
\newcommand{\Planck}{\emph{Planck}}
\newcommand{\lcdm}{$\Lambda$CDM}
\newcommand{\mH}{m_{\nu_H}} 
\newcommand{\ml}{m_{\nu_L}} 
\newcommand{\ev}{\mathrm{eV}}

\newcommand{\class}{\texttt{CLASS}} 
\newcommand{\connect}{\texttt{CONNECT}}

\renewcommand{\arraystretch}{1.2}

\newcommand{\appref}[1]{\hyperref[#1]{App.~\ref*{#1}}}

\makeatletter  
\renewcommand\onecolumngrid{
\do@columngrid{one}{\@ne}
\def\set@footnotewidth{\onecolumngrid}
\def\footnoterule{\kern-6pt\hrule width 1.5in\kern6pt}
}
\renewcommand\twocolumngrid{
        \def\footnoterule{
        \dimen@\skip\footins\divide\dimen@\thr@@
        \kern-\dimen@\hrule width.5in\kern\dimen@}
        \do@columngrid{mlt}{\tw@}
}
\makeatother

\begin{document}

\preprint{APS/123-QED}

\title{Neutrino decays as a natural explanation of the neutrino mass tension} 

\author{Guillermo Franco Abellán}
\email{g.francoabellan@ific.uv.es}
\affiliation{
Instituto de Física Corpuscular (IFIC), CSIC-Universitat de València, \\ Parc Científic UV, c/ Catedrático José Beltrán, 2, E-46980 Paterna (València), Spain}
\affiliation{GRAPPA Institute, Institute for Theoretical Physics Amsterdam, \\University of Amsterdam, Science Park 904, 1098 XH Amsterdam, The Netherlands}

\date{\today}

\begin{abstract} 
A new tension is emerging between the tight cosmological upper bounds on the total neutrino mass ($\sum m_\nu \lesssim 0.06 \, \ev$) and the lower limits from oscillation experiments, with potentially far-reaching implications for cosmology and particle physics. Neutrinos decaying into massless BSM particles with lifetimes $\tau_\nu \sim 0.01-1\, \rm{Gyr}$ represent a theoretically well-motivated mechanism to reconcile such measurements. Using DESI DR2 and CMB datasets, we show that such invisible decays relax the bound on the total neutrino mass up to $\sum m_\nu < 0.23 \, \ev$ (95\%), restoring full agreement with oscillation data. We also present the first late-time cosmological analysis of invisible neutrino decays into lighter neutrinos in a manner consistent with the measured mass splittings. In contrast to the decays into massless BSM particles, we find that this scenario only marginally alleviates---or even tightens---the cosmological neutrino mass bounds, depending on the mass ordering.
\end{abstract}

\maketitle

\section{\label{sec:introduction}Introduction}

Neutrino oscillation experiments have firmly established that at least two of the three neutrino species possess non-zero masses, providing the only clear evidence to date of physics beyond the Standard Model (SM). In particular, these experiments have precisely measured two independent mass splittings: $|\Delta m^2_{31}| \equiv |m_3^2-m_1^2| \simeq 2.5 \times 10^{-3} \ev^2$ and $\Delta m^2_{21} \equiv m_2^2-m_1^2 \simeq 7.5 \times 10^{-5} \ev^2$ \cite{deSalas:2020pgw}. However, the absolute neutrino mass scale---parameterized either by $\sum m_\nu = m_1 + m_2 + m_3$ or $m_{\rm lightest}$---remains unknown, as does the neutrino mass ordering, which could be normal (NO; $m_3 \gg m_2 > m_1$) or inverted (IO; $m_2 > m_1 \gg m_3$). Assuming $m_{\rm lightest} \rightarrow 0$, the measured mass splittings imply a lower bound on the total neutrino mass of $\sum m_\nu > 0.059 \, \ev$ for NO and $\sum m_\nu > 0.098 \, \ev$ for IO \cite{Esteban:2024eli}. Explaining the origin of neutrino masses remains a major open question in fundamental physics. 

On the other hand, in recent years cosmological observations have placed increasingly stringent constraints on $\sum m_\nu$. Within $\Lambda$CDM, the latest upper limit reported by the DESI collaboration is $\sum m_\nu < 0.064 \, \ev~(95\%)$ \cite{DESI:2025zgx}, obtained from the combination of their baryon acoustic oscillation (BAO) data with \Planck~PR4 Cosmic Microwave Background (CMB) likelihoods. This bound is over twenty times stronger than the direct laboratory limit $\sum m_\nu <1.35 \, \ev~(90\%)$ from KATRIN \cite{KATRIN:2024cdt}, and is remarkably close to the minimum value allowed by neutrino oscillation experiments. Intriguingly, when allowing for effective negative masses, the cosmological posterior peaks in the negative region \cite{Craig:2024tky, Naredo-Tuero:2024sgf}, and the tension with oscillation measurements reaches $3\sigma$ \cite{Elbers:2024sha,Elbers:2025vlz}. These unphysically small limits on $\sum m_\nu$ in CMB+BAO data originate from the BAO preference for a low $\Omega_m$ \cite{Loverde:2024nfi,Jhaveri:2025neg,Lynch:2025ine}, and can be relaxed by introducing dynamical dark energy \cite{Elbers:2024sha, DESI:2025ffm}, modified recombination \cite{Baryakhtar:2024rky,Lynch:2024hzh}, a large reionization optical depth $\tau_{\rm reio}$ \cite{Jhaveri:2025neg,Sailer:2025lxj, Tan:2025obi}, a suppressed growth rate \cite{Giare:2025ath} or an excess of CMB lensing \cite{Cozzumbo:2025ewt}.

Alternatively, the neutrino mass tension could point to new physics in neutrino sector, such as time dependent $m_\nu$ \cite{Lorenz:2018fzb, Lorenz:2021alz,Ghedini:2025epp}, long-range forces \cite{Esteban:2021ozz, Esteban:2022rjk}, or other forms of non-standard neutrino interactions \cite{Beacom:2004yd,Cuoco:2005qr,Farzan:2015pca,Kreisch:2019yzn,Oldengott:2019lke,Hooper:2021rjc,Alvey:2021sji,Escudero:2022gez,Poudou:2025qcx}, including neutrino decays \cite{Serpico:2007pt,Serpico:2008zza}. The possibility of unstable neutrinos was already considered in the 70's \cite{Bahcall:1972my}, and in fact, two neutrino mass eigenstates decay within the SM, albeit with lifetimes vastly longer than the age of the Universe, $\tau_\nu > 10^{23}\, t_{\rm U}$. Many SM extensions, however, predict substantially shorter neutrino lifetimes, e.g., \cite{Chikashige:1980ui,Schechter:1981cv,Gelmini:1980re,Gelmini:1983ea,Georgi:1990se,Burgess:1992dt,Joshipura:1992vn,Davidson:2005cs,Dvali:2016uhn}. If neutrinos decay on timescales $\tau_\nu \sim 0.01 - 0.1\,t_{\rm U}$, it has been shown that the neutrino mass bounds from cosmology can be significantly relaxed  \cite{Chacko:2019nej}. This generally requires invisible decay channels, since radiative decays are strongly constrained,  $\tau_\nu > (10^2-10^4)t_{\rm U}$ \cite{Mirizzi:2007jd,Aalberts:2018obr}. So far, all cosmological analyses of late-time invisible neutrino decays have assumed that the decay products are massless BSM particles, i.e., dark radiation (DR). Specifically, the analysis performed in \cite{FrancoAbellan:2021hdb}, which includes \Planck~PR3 + SDSS BAO data, showed that in this regime the neutrino mass bound is relaxed up to $\sum m_\nu < 0.42 \, \ev \, (95\%)$. 

In this paper, we perform a thorough assessment of invisible neutrino decays in light of the current neutrino mass tension. We first update the neutrino mass constraints for late-time decays into DR using the latest DESI DR2 BAO measurements. Although the resulting bounds on $\sum m_\nu$ are improved by a factor of two compared to those derived from SDSS BAO, they can still restore full consistency with neutrino oscillation results. In addition, we present the first late-time cosmological analysis of decays into final states containing active neutrinos. This scenario is very appealing from a model building perspective, as it can accommodate a realistic neutrino mass spectrum. We uncover new cosmological signatures and find that, depending on the mass ordering, such decays either leave the neutrino mass bound only marginally alleviated or even tighten it.

\section{\label{sec:theory}Theory}

Invisible neutrino decays naturally arise in \emph{Majoron} models, where the spontaneous breakdown of global lepton number generates light neutrino masses and leads to a massless Goldstone boson that couples to neutrinos \cite{Chikashige:1980ui,Schechter:1981cv,Gelmini:1980re}. Here, we assume the following Lagrangian 
\begin{equation}
\mathcal{L}_{\rm int} = \mathfrak{g}\,\bar{\nu}_H \nu_{L}\phi,   
\end{equation}
describing effective Yukawa interactions between a heavier active neutrino $\nu_H$, a lighter neutrino $\nu_L$, and a massless scalar $\phi$. The corresponding rate of neutrino decay $\nu_{H} \rightarrow \nu_L +\phi$ is\footnote{Three-body neutrino decays can become relevant when medi-
ated by Majorons with a finite sub-MeV mass; however, such couplings are strongly constrained by the CMB  \cite{Escudero:2019gvw,Escudero:2020ped}.}
\begin{equation}
\Gamma_\nu = \frac{\mathfrak{g}^2}{4 \pi} \frac{\left(\mH-\ml \right)\left(\mH + \ml \right)^3}{ \mH^3},
\label{eq:decay_rate}
\end{equation}
where it is assumed that neutrinos are Majorana particles. In what follows, we consider two main scenarios depending on the nature of the lighter neutrino $\nu_L$. \ 

\emph{Scenario A (decays into DR)}. In this case, $\nu_L$ is taken to be a new massless sterile neutrino state $\nu_4$, such that the decay products $\{\nu_4,\phi\}$ behave as a DR fluid. These decays can be realized, for example, in minimal neutrino mass models within a $U(1)_{\mu -\tau}$ flavor symmetry \cite{Escudero:2020ped}. Following previous studies of decays of the type $\nu_{H} \rightarrow \rm{DR}$ \cite{Chacko:2019nej,FrancoAbellan:2021hdb}, we assume all three neutrinos to be degenerate in mass ($\sum m_\nu =3\mH$), which provides a reasonable approximation for current and future cosmological data \cite{Archidiacono:2020dvx,Herold:2024nvk}. From \autoref{eq:decay_rate}, this implies that all three neutrinos share the same decay rate $\Gamma_\nu$. \  

\emph{Scenario B (decays into lighter SM neutrinos)}.  In this case $\nu_L$ is  an active neutrino, so the decay mass gap corresponds to the measured squared mass differences, and only $\phi$ plays the role of the DR. Generally, one expects non-zero decay rates $\Gamma_{i\rightarrow j}$ between all possible pairs in the process $\nu_i \rightarrow \nu_j + \phi$. Nevertheless, from the measured mass splittings, the decay rates follow to a good approximation $\Gamma_{3\rightarrow 2} \simeq \Gamma_{3\rightarrow 1} \gg \Gamma_{2\rightarrow 1}$ (NO) and $\Gamma_{2\rightarrow 3} \simeq \Gamma_{1\rightarrow 3} \gg \Gamma_{2\rightarrow 1}$ (IO), meaning that $\nu_1$ and $\nu_2$ can be treated as degenerate species and the three-state system can effectively be reduced to a two-state one \cite{Barenboim:2020vrr,Chen:2022idm}. We henceforth adopt this approximation and consider two decay channels separated by one common atmospheric mass gap $|\Delta m_{31}^2|$. We distinguish two cases according to the mass ordering: B1 (NO; $\nu_{3} \rightarrow \nu_{1,2}+\phi$) and B2 (IO; $\nu_{1,2} \rightarrow \nu_3+\phi$).\footnote{In the B1(B2) case, all collision terms in the Boltzmann equations for $\nu_H$ ($\nu_L$) and $\phi$ are multiplied by 2, as are all momentum-integrated quantities associated with $\nu_L$ ($\nu_H$).}  

\begin{figure*}[ht!]
\centering
\includegraphics[scale=0.42]{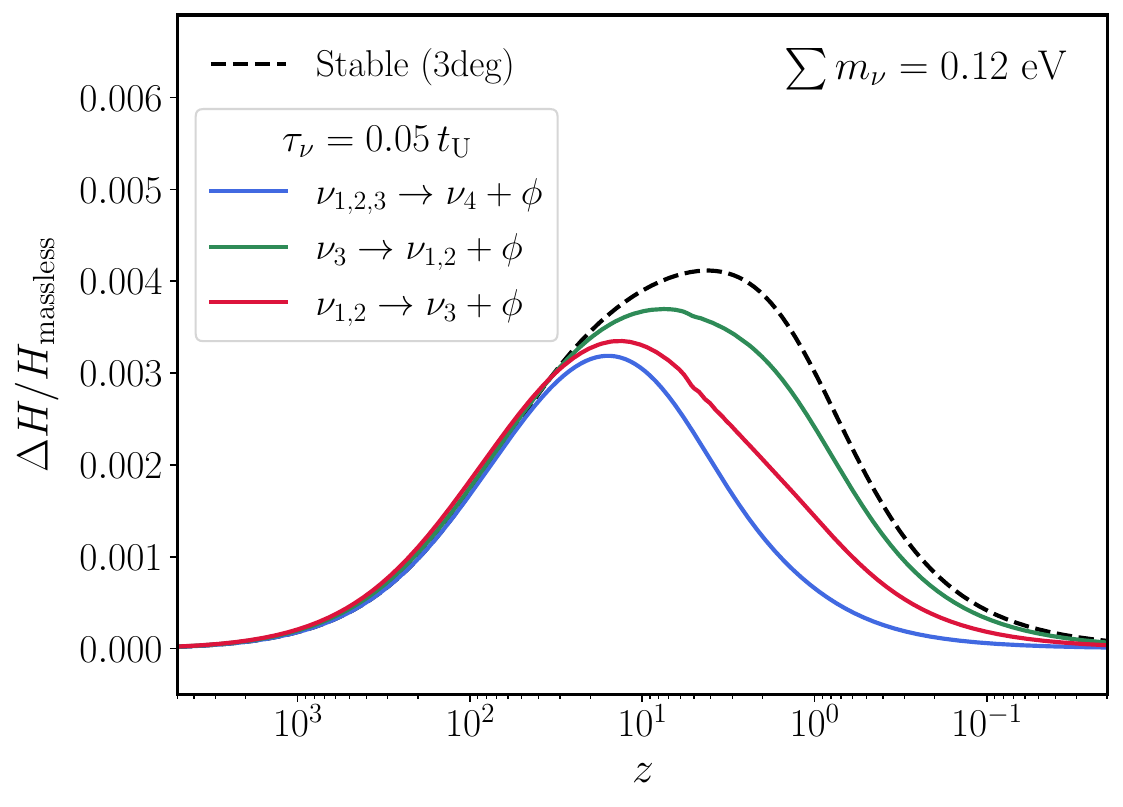}
\hspace{5mm}
\includegraphics[scale=0.42]{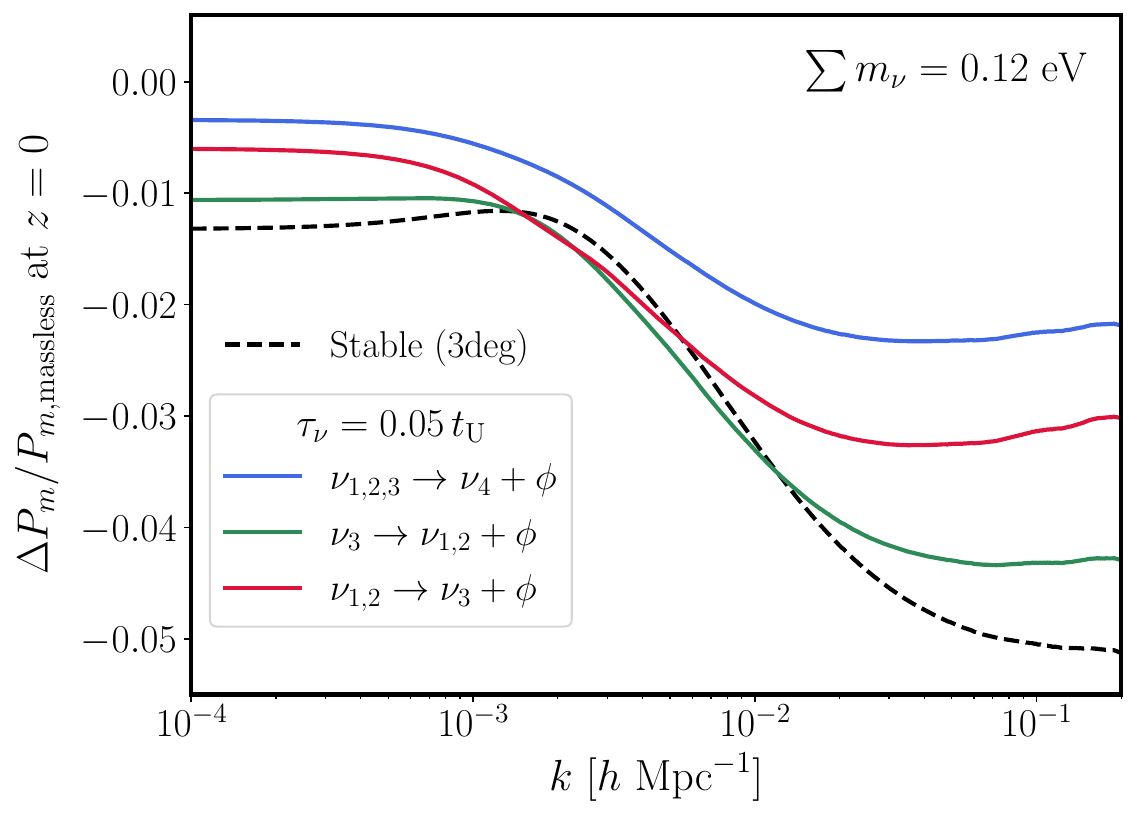}
\caption{Effects of massive neutrinos on the Hubble rate (\emph{left panel}) and on the present-day linear matter power spectrum (\emph{right panel}), relative to the massless case, for fixed cosmological parameters $\{H_0, \omega_{\rm b}, \omega_{\rm cdm}, A_s, n_s, \tau_{\rm reio}, N_{\rm eff}\}$ and total neutrino mass $\sum m_\nu = 0.12 \, \ev$. We compare the case of three stable degenerate neutrinos (dashed lines) to the decay scenarios A, B1, and B2 introduced in the main text (solid lines), all assuming a lifetime equal to 5\% of the age of the Universe.}
\label{fig:Hz_Pk_residuals}
\end{figure*}

For both scenarios A and B, we assume that the population of DR particles is produced at late times only via neutrino decays. We further assume that neutrinos decay after becoming non-relativistic. Hence, inverse decay processes are kinematically suppressed and can be neglected. The regime of non-relativistic decays  is set by the threshold condition on the neutrino lifetime $\tau_\nu > H^{-1} (z_{\rm nr})$, where $H(z_{\rm nr})$ is given by \cite{FrancoAbellan:2021hdb}
\begin{equation}
H (z_{\rm nr} ) =H_0 \sqrt{\Omega_m}\left(\frac{m_{\nu_H}}{3T_{\nu 0}}\right)^{3 / 2}.   
\end{equation}
For the B1 case, the lifetime is computed as $\tau_\nu = 1/(2\Gamma_\nu)$, to account for its two distinct decay modes. Invisible neutrino decays in the ultra-relativistic limit (i.e., occurring before recombination) are also possible, but they lead to a completely different phenomenology, e.g., \cite{Basboll:2008fx,Archidiacono:2013dua,Escudero:2019gfk,Barenboim:2020vrr,Chen:2022idm}.

For scenario A, we use a version of the Einstein–Boltzmann solver \class~\cite{Lesgourgues:2011re,Blas:2011rf}, extended in \cite{Holm:2022eqq} to model warm species decaying into DR. For scenario B, we use our modification of this code that accounts for the effects of a finite $\ml$. The background and inhomogeneous Boltzmann equations governing the cosmological evolution of the three species ${\nu_H, \nu_L, \phi}$ are given in \cite{Terzaghi:2025pvb}; they correspond to the non-relativistic limit of the full system derived in \cite{Barenboim:2020vrr} for the processes $\nu_H \leftrightarrow \nu_L + \phi$. 

Massive neutrinos influence cosmological observables (e.g., CMB, BAO) primarily through their impact on the expansion history and the growth of structure \cite{Lesgourgues:2006nd,Bertolez-Martinez:2024wez}. As depicted in \autoref{fig:Hz_Pk_residuals}, late-time invisible decays modify this impact and can therefore alter neutrino mass bounds. In the stable case, massive neutrinos produce two well-known effects: i) a boost in the Hubble rate $H(z)$ during the matter era, as they contribute to $\Omega_m$ at late times, and ii) a step-like suppression in the matter power spectrum $P_m(k)$, since they add to Hubble friction but they do not cluster on scales $k > k_{\rm nr}$, where $k_{\rm nr}$ denotes their minimum free-streaming wavenumber $k_{\rm nr} = k_{\rm fs}(z_{\rm nr})$ \cite{Shoji:2010hm}. In decay scenario A, all neutrino mass energy is transferred to radiation around $t \sim \tau_\nu$, substantially weakening both of these effects. In contrast, in decay scenario B the decay products always contain a lighter massive neutrino, so these effects are weakened to a lesser extent. Interestingly, such decay scenarios predict a distinctive signature in $P_m(k)$: a reduction in $k_{\rm nr}$. This arises because the decays produce a long high-energy tail in the phase-space of $\nu_L$, raising its average momentum and delaying its transition to the non-relativistic regime (see \appref{app:PSD_and_kFS}). The resulting shift in $k_{\rm nr}$ is more prominent in B2 than in B1, since the phase-space of $\nu_L$ is populated by two $\nu_H$ particles rather than just one. To the best of our knowledge, this cosmological feature of the decays $\nu_i \rightarrow \nu_j + \phi$ had never been pointed out before, and as we will see it leads to neutrino mass bounds that are radically different from the ones of the decays  $\nu_i \rightarrow \nu_4 + \phi$. \

\section{\label{sec:methodos}Methods}

Cosmological inference for invisible neutrino decays can be very time consuming, particularly for decays into lighter SM neutrinos. Indeed, computing the dynamics of the $\nu_H$ and $\nu_L$ species necessitates tracking their full phase-space evolution, with the required number of momentum bins and maximum momentum varying significantly across the decay parameter space \cite{Terzaghi:2025pvb}. As a consequence, solving the linearised Einstein-Boltzmann system can take up to tens of CPU core-hours for certain decay configurations, rendering MCMC analyses computationally prohibitive. To overcome this, we make use of neural network emulators built with \connect~\cite{Nygaard:2022wri,Nygaard:2024lna}, to emulate the output of our modified \class~version for both decay scenarios A and B (further details given in \appref{app:emulators}).\

Our baseline dataset combines the BAO distance measurements from DESI DR2 \cite{DESI:2025zgx} with the TT,EE,TE CMB power spectra from \Planck, specifically using the \texttt{Commander}, \texttt{SimAll} (for $\ell < 30$) and \texttt{Plik} (for $\ell \geq 30$) likelihoods, along with the CMB lensing reconstruction from the PR3 data release \cite{Planck:2019nip,Planck:2018lbu}. We do not include any supernovae (SNe) datasets, as their influence on neutrino mass constraints has been shown to be very weak in the presence of a cosmological constant \cite{DESI:2025zgx}. Our baseline constraints also omit the latest \Planck~PR4 likelihoods \cite{Rosenberg:2022sdy,Carron:2022eyg} and the CMB lensing data from ACT DR6 \cite{ACT:2023dou,ACT:2023kun}, since incorporating them would require to model non-linear corrections and hence to recalibrate the \texttt{HaloFit} \cite{Takahashi:2012em} or \texttt{HMcode} \cite{Mead:2020vgs} algorithms for decaying neutrino cosmologies. In \appref{app:CMB_likelihoods}, we examine the impact of including these newer CMB likelihoods and using \texttt{HaloFit} for the decay scenario A.\

We perform MCMC explorations of the parameter space with \texttt{MontePython-v3} \cite{Audren:2012wb, Brinckmann:2018cvx}, interfaced with our \class-based emulators. For all runs, we assume flat priors on $\{\omega_b, \omega_{\rm cdm}, H_0, n_s, \ln(10^{10}A_s), \tau_{\rm reio}\}$. In addition, we impose flat priors on $\mH \in [0, 0.6] \, \ev$ and $\log_{10}(\tau_\nu/t_{\rm U}) \in [-3.55,-0.15]$ for scenario A, and on $\ml \in [0, 0.6] \, \ev$ and $\log_{10}(\tau_\nu/t_{\rm U}) \in [-3.55, -0.95]$ for scenario B. To ensure a self-consistent treatment, we exclude the regime of relativistic decays from the scan by imposing the prior $\tau_\nu > H^{-1}(z_{\rm nr})$. We perform additional runs for the stable limit of each decay scenario, i.e., the one with the same neutrino mass spectrum and $\Gamma_\nu =0$. We note that, under the degenerate-$\nu_{1,2}$ approximation, the total neutrino mass is  $\sum m_\nu  = 2\ml + \mH$ for B1 and $\sum m_\nu  = \ml + 2\mH $ for B2, where $\mH^ 2 =  \ml^ 2+|\Delta m_{31}^2|  $. These relationships enforce lower limits of $\sum m_\nu >0.05 \, \ev$ and $\sum m_\nu >0.1 \, \ev$ for B1 and B2, respectively. We ran 16 chains until a Gelman-Rubin \cite{Gelman:1992zz} $R-1 <0.02$ was obtained, and computed the posterior distributions with \texttt{GetDist} \cite{Lewis:2019xzd}.\

\section{\label{sec:results}Results}

\subsection{Decays into dark radiation}

\begin{figure}[h!]
\centering
\includegraphics[width=0.9\linewidth]{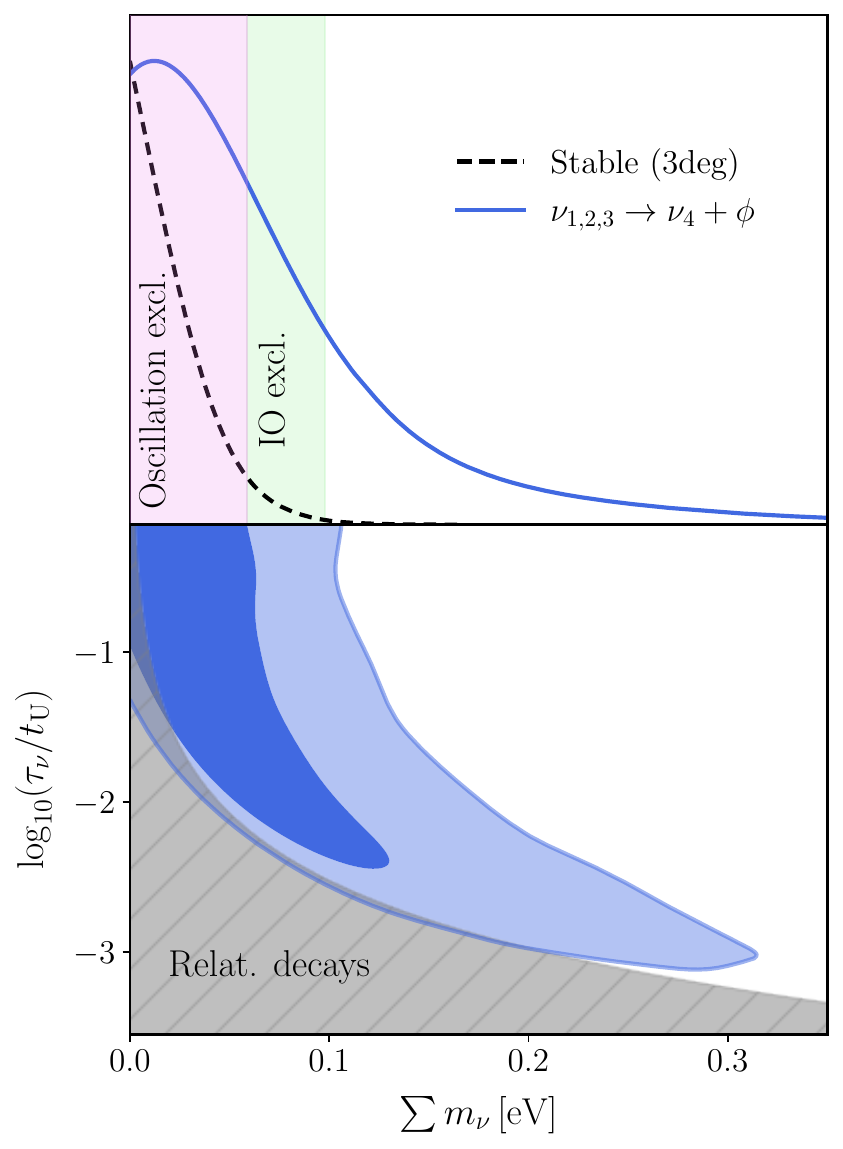}
\caption{Constraints from DESI DR2 BAO and \Planck~PR3 CMB on the total neutrino mass $\sum m_\nu$ and neutrino lifetime $\log_{10}(\tau_\nu/t_{\rm U})$, assuming that three degenerate neutrinos decay into a $\nu_4+\phi$ dark radiation fluid (solid blue). The stable limit is denoted by the black dashed line.  The pink and green shaded regions indicates exclusion from neutrino oscillation experiments, while the scratched area indicates the regime of relativistic decays excluded from the prior.}
\label{fig:decay_DR}
\end{figure}

The main result of our analysis for decays into DR is presented in  \autoref{fig:decay_DR}. There is a large negative correlation between the total neutrino mass and neutrino lifetime, such that values as large as $\sum m_\nu \sim 0.2-0.3 \, \ev$ remain consistent with cosmological data if neutrinos decay into DR with lifetimes $\tau_\nu \sim 0.001-0.01\,t_{\rm U}$. Consequently, the marginalized 95\% CL bound on the total neutrino mass is relaxed from $\sum m_\nu < 0.062 \, \ev$ in the stable case to $\sum m_\nu < 0.23 \, \ev$ when such decays are allowed. This updated bound is roughly a factor of two stronger than the limit on the same scenario derived with \Planck~PR3 and SDSS BAO data \cite{FrancoAbellan:2021hdb}. We have verified that the improvement is driven primarily by: i) the increased constraining power of DESI\footnote{Even if we explore decays happening before the period relevant for BAO observables, the addition of DESI BAO data tightens the constraints on parameters degenerate with $\sum m_\nu$ (e.g., $H_0, \,\Omega_m$), which results in a stronger neutrino mass bound.} and, to a lesser extent,  ii) the use of a prior $\sum m_\nu >0$ instead of $\sum m_\nu >0.06 \, \ev$ as adopted in \cite{FrancoAbellan:2021hdb}. Nevertheless, this still represents a substantial relaxation (by almost a factor of four) in comparison to the stable neutrino case, removing any tension with laboratory results. To quantify this, we use a goodness-of-fit-loss metric
based on the $\Delta \chi^2_{\rm min}$, evaluated at the best-fit points\footnote{We obtain the best-fit points by minimising the $\chi^2$ with the method discussed in Appendix D of \cite{Schoneberg:2021qvd}.} with $\sum m_\nu$ free and
with $\sum m_\nu$ fixed to the lower bounds from oscillation experiments. As detailed in \autoref{tab:tension_level}, the tension with the NO case ($\sum m_\nu > 0.059~\ev$) is reduced from $\sim 2.1\sigma$ for stable neutrinos to $\sim 1.3\sigma$ for decays to DR. In the IO case ($\sum m_\nu > 0.098~\ev$), the tension is reduced from $\sim 3\sigma$ to $\sim 1.5\sigma$. It is important to stress that, despite the significant broadening of the $\sum m_\nu$ posterior,  the decay scenario does not yield a positive detection of neutrino masses. In both the stable and unstable cases, the best-fit value of  $\sum m_\nu$ lies at the edge of the prior ($\sum m_\nu = 0$), and the decays to DR provide only a marginal improvement in the fit, with $\Delta \chi^2_{\rm min} = -1.7$ relative to the stable limit.

\begin{table}[t]
\centering
\renewcommand{\arraystretch}{1.4}
\begin{tabular}{|c|cc|cc|}
\hline
\multirow{2}{*}{Model} & \multicolumn{2}{c|}{Tension with NO} & \multicolumn{2}{c|}{Tension with IO} \\ 
 & $\Delta \chi^2_{\rm min}$ & $Q$ & $\Delta \chi^2_{\rm min}$ & $Q$ \\ 
\hline
Stable (3deg) & $-4.3$  & $2.1\sigma$ & $-9.1$  & $3.0\sigma$ \\ 
$\nu_{1,2,3} \rightarrow \nu_4+\phi$  & $-1.7$  & $1.3\sigma$ & $-2.4$  & $1.5\sigma$  \\ 
\hline
\end{tabular}
\caption{The tension $Q$ between the cosmological (\Planck~PR3 + DESI DR2 BAO) and terrestrial constraints on $\sum m_\nu$, for stable neutrinos and neutrino decays into DR, assuming three degenerate mass states. The tension is estimated by evaluating $\Delta \chi^2 = \chi^2 (\sum m_\nu~\mathrm{free})-\chi^2 (\sum m_\nu=x)$ at the best-fit points (with $x=0.059 \, \rm{eV}$ for NO and $x=0.098 \, \rm{eV}$ for IO) and converting it to a Gaussian $N\sigma$ level.}
\label{tab:tension_level}
\end{table}

From a particle physics perspective, the neutrino lifetimes required to relax the mass bounds up to $\sum m_\nu \sim 0.2-0.3~\ev$ correspond to neutrino-Majoron couplings of order $\mathfrak{g} \sim 10^{-14}$. Data from SN1987 \cite{Kachelriess:2000qc,Farzan:2002wx} and Big Bang Nucleosynthesis \cite{Escudero:2019gfk} exclude Majoron couplings with active neutrinos in the range $\mathfrak{g} \gtrsim 10^{-7}-10^{-5}$, while CMB constraints on neutrino free-streaming rule out $\mathfrak{g} \gtrsim 10^{-11}-10^{-9}$ \cite{Barenboim:2020vrr}. All of these bounds disfavor neutrino lifetimes that are orders of magnitude shorter than the ones considered here. In the context of neutrino mass models based on a spontaneously broken $U(1)_{\mu-\tau}$ symmetry, effective couplings $\mathfrak{g} \sim 10^{-14}$ and neutrino masses $\sum m_\nu \sim 0.2-0.3~\ev$ are associated with $M_R/y^2 = m_\nu/\mathfrak{g}^2 \sim 10^{17}-10^{18}\, \rm{GeV}$, where $M_R$ is the mass of the heavy right-handed neutrino and $y$ is the Yukawa coupling of the UV complete theory \cite{Escudero:2020ped}. This naturally points to GUT-scale seesaw for $y\sim 0.1$. We deduce that the parameter space shown in \autoref{fig:decay_DR} can be consistently embedded within minimal neutrino mass models proposed in the literature and is compatible with current observational constraints.

\subsection{Decays into lighter neutrinos}

For decays into lighter SM neutrinos, one expects a much weaker relaxation of the mass bounds. By fermion number conservation, the effective sum of neutrino masses inferred from the present neutrino energy density $\rho_{\nu,0}$ is given by $\sum m_\nu^{\rm cosmo} = 3\ml$, assuming all neutrinos except the lightest one have decayed. Since the boost in $H(z)$ and the high-$k$ constant suppression in $P_m(k)$ (see \autoref{fig:Hz_Pk_residuals}) are both controlled by $f_\nu \equiv \rho_{\nu,0}/\rho_{\rm m,0}$, the relaxation of the bound on $\sum m_\nu$ can be estimated as a shift $\sum m_\nu -3\ml$ \cite{Escudero:2020ped}. This implies a maximum alleviation of $0.05 \, \ev$ ($0.1 \, \ev$) for scenarios B1 (B2) in the limit $\ml \rightarrow 0$. In practice, this naïve estimate can be altered if neutrinos in the final states are relativistic and/or if the decays produce additional effects which cannot be captured solely by a change in $\rho_{\nu,0}$. \ 

\begin{figure}[h!]
\centering
\includegraphics[width=0.9\linewidth]{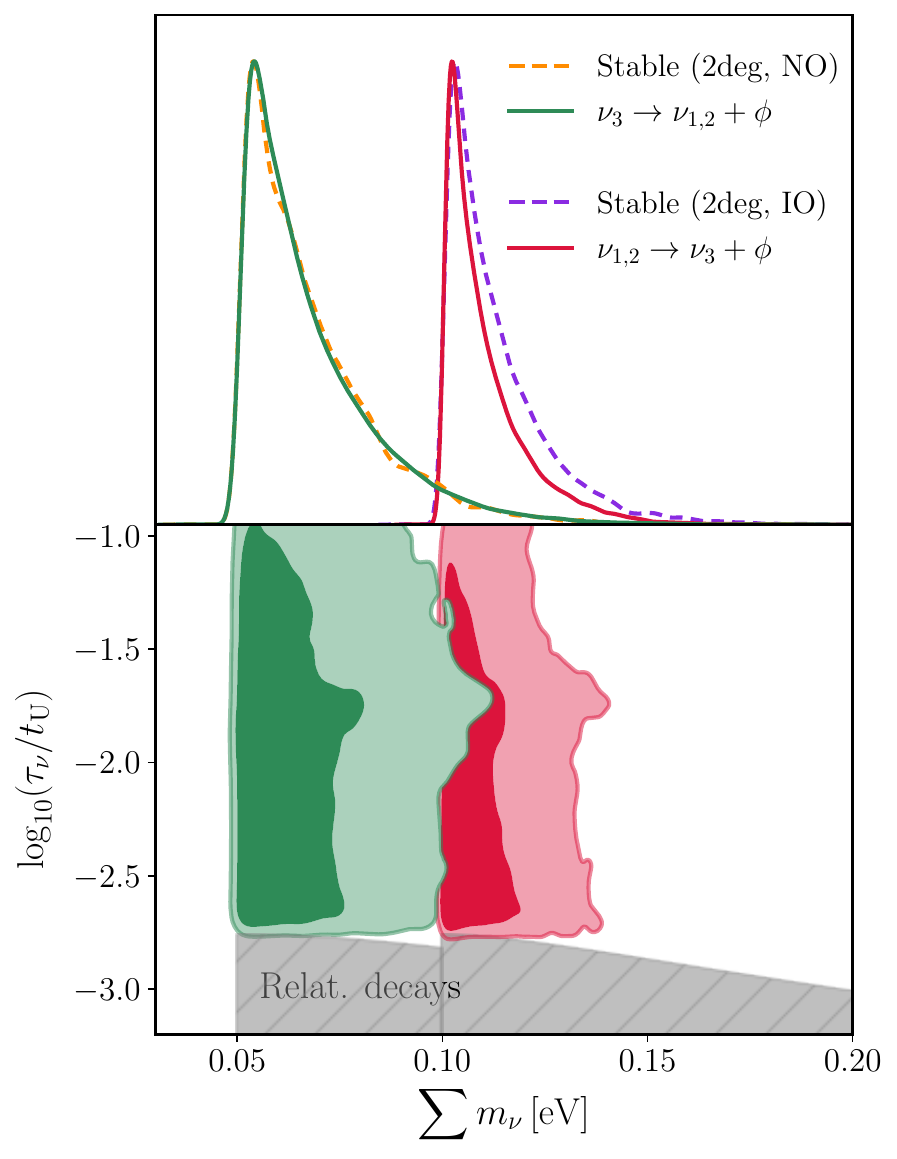}
\caption{Constraints from DESI DR2 BAO and \Planck~PR3 CMB on the total neutrino mass $\sum m_\nu$ and neutrino lifetime $\log_{10}(\tau_\nu/t_{\rm U})$, assuming that neutrinos decay with the atmospheric mass gap in the degenerate-$\nu_{1,2}$ approximation, for the normal (solid green) and inverted (solid red)  orderings. The yellow and purple dashed lines denote the corresponding stable limits. The scratched areas indicate the regime of relativistic decays excluded from the prior.}
\label{fig:decay_NCDM}
\end{figure}

\autoref{fig:decay_NCDM} shows that the neutrino mass constraints are dominated by the oscillation lower bounds, and the small shifts in $\sum m_\nu$ induced by the decays differ from the simple expectation above. Quantitatively, scenario B1 yields a negligible relaxation (stable: $\sum m_\nu < 0.1 \, \ev$ vs. decay: $\sum m_\nu < 0.102 \, \ev$), whereas scenario B2 actually yields a \emph{tightening} of the mass bound (stable: $\sum m_\nu < 0.139 \, \ev$ vs. decay: $\sum m_\nu < 0.132 \, \ev$), with all limits at 95\% CL. These constraints result from the interplay of two competing effects: i) a decrease in the boost of $H(z)$ and in the plateau-like suppression of $P_{\rm m}(k)$, which weakens the imprint of neutrino masses on CMB and BAO observables, and ii) a decrease in the cutoff wavenumber $k_{\rm nr}$, which enhances the sensitivity of the CMB lensing potential to neutrino masses, particularly for lifetimes $\tau_\nu \sim (10^{-1.5}$–$10^{-1})t_{\rm U}$.\footnote{This implies that the stable neutrino limit in scenario B is only recovered for $\tau_\nu \gg t_{\rm U}$, rather than $\tau_\nu \sim t_{\rm U}$ as in scenario A.} In the B1 case, the two effects nearly cancel, leading to an almost unchanged bound, while in the B2 case the reduction in $k_{\rm nr}$ dominates, leading to a tightening of the bound. These non-trivial effects demonstrate that a full Boltzmann treatment is essential for a reliable analysis of the decays $\nu_i \rightarrow \nu_j +\phi$.

\section{\label{sec:discussions}Conclusions}

Invisible neutrino decays provide a compelling framework to explain the non-detection of neutrino masses in cosmological observations. Unlike many exotic models proposed to relax the cosmological neutrino mass bounds, this framework possesses a concrete Lagrangian formulation, naturally connected with neutrino mass generation mechanisms. Using DESI DR2 BAO + \Planck~PR3 data, we have shown that non-relativistic neutrino decays into DR ($\nu_i \rightarrow \nu_4+\phi$) relax the 95\% CL limit on the total neutrino mass from $\sum m_\nu < 0.062  \, \ev$ to $\sum m_\nu < 0.23  \, \ev$. This completely eliminates any tension with the oscillation lower bounds in both the normal and inverted orderings. We have also found that neutrino decays into lighter SM neutrinos ($\nu_i \rightarrow \nu_j+\phi$), while allowing to derive mass-spectrum–consistent cosmological constraints, yield essentially no relaxation or even a tightening of the $\sum m_\nu$ bound relative to the stable case, depending on the mass ordering.\ 

These late-time invisible decay channels leave characteristic imprints on the time evolution of the matter power suppression and, for decays into lighter SM neutrinos, on the associated cutoff scale. Hence, upcoming tomographic weak-lensing surveys such as \Euclid~\cite{Euclid:2024imf} or LSST \cite{LSST:2008ijt} may enable for an independent determination of neutrino mass and lifetime, or improve existing lifetime bounds by several orders of magnitude \cite{Chacko:2020hmh}. This motivates the development of methods for modeling non-linear scales in cosmologies with decaying neutrinos. It will also be interesting to extend current analysis to the regime of semi-relativistic decays, to asses whether neutrino decays could accommodate a potential signal at KATRIN \cite{KATRIN:2024cdt}. Looking further ahead, a direct detection of the cosmic neutrino background (C$\nu$B) by future experiments such as PTOLEMY \cite{PTOLEMY:2018jst,PTOLEMY:2019hkd} would have important implications for invisible neutrino decays. Crucially, a direct C$\nu$B detection would rule out the possibility that neutrinos have fully decayed into DR at some stage during cosmic history, or strongly constrain decays between neutrino mass eigenstates with lifetimes of the order $\tau_\nu \sim t_{\rm U}$ \cite{Akita:2021hqn,Terzaghi:2025pvb}.

\begin{acknowledgments}

The author would like to thank Nicola Terzaghi, Shin'ichiro Ando and Vivian Poulin for interesting discussions, as well as Andreas Nygaard for useful comments on the \connect~emulator and Thejs Brinckmann for kindly sharing the ACT DR6 CMB lensing and \Planck~PR4 likelihoods for \texttt{MontePython-v3}. The author acknowledges support from the European Research Council (ERC) under the European Union's Horizon 2020 research and innovation programme (Grant agreement No. 864035 - Undark). The main analysis for this work was carried out on the Snellius Computing Cluster at SURFsara. 
\end{acknowledgments}

\section*{Data availability}

Our modified \class~version is publicly available at \cite{francoabellan2024classnudecay}. 
\hspace{1mm}The data underlying this article will
be shared on reasonable request to the corresponding author. 

\appendix

\section{Free-streaming scale of the lighter neutrino}\label{app:PSD_and_kFS}

\begin{figure*}[ht!]
\centering
\includegraphics[scale=0.44]{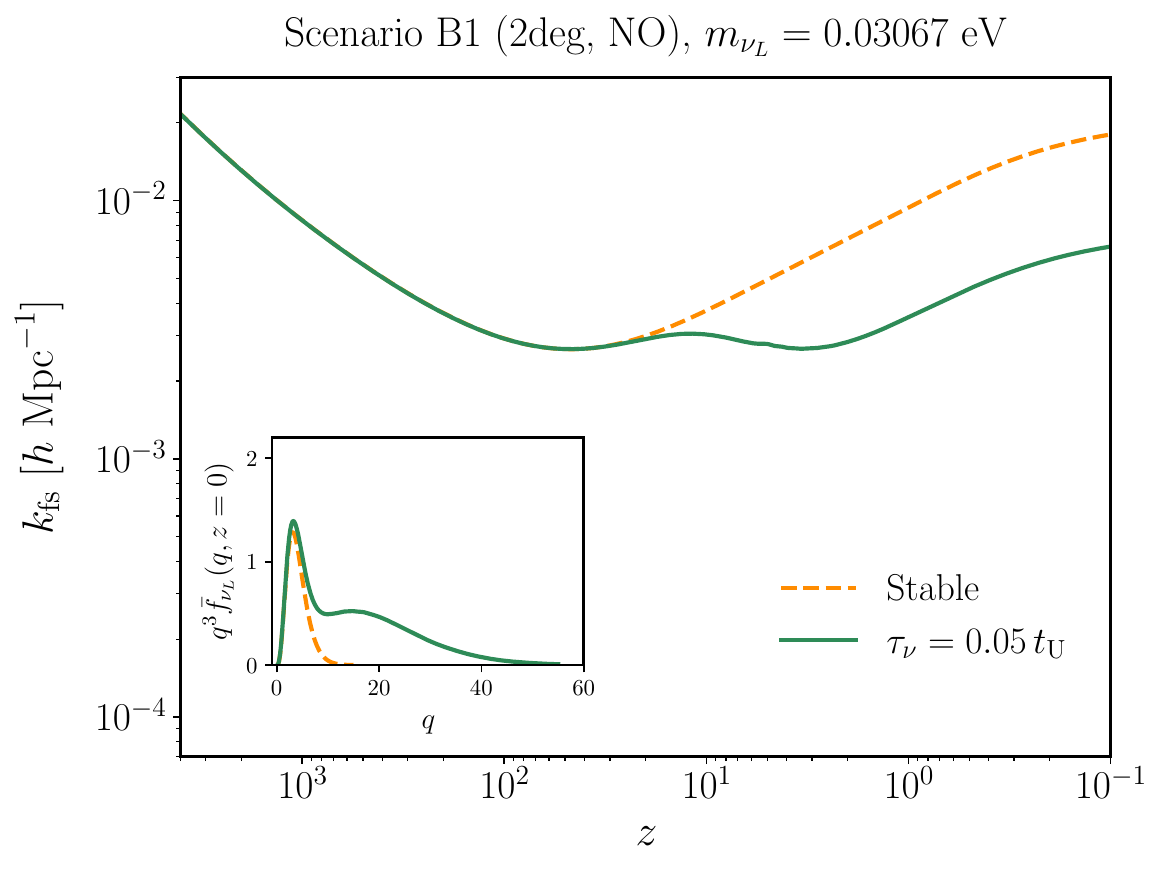}
\hspace{2mm}
\includegraphics[scale=0.44]{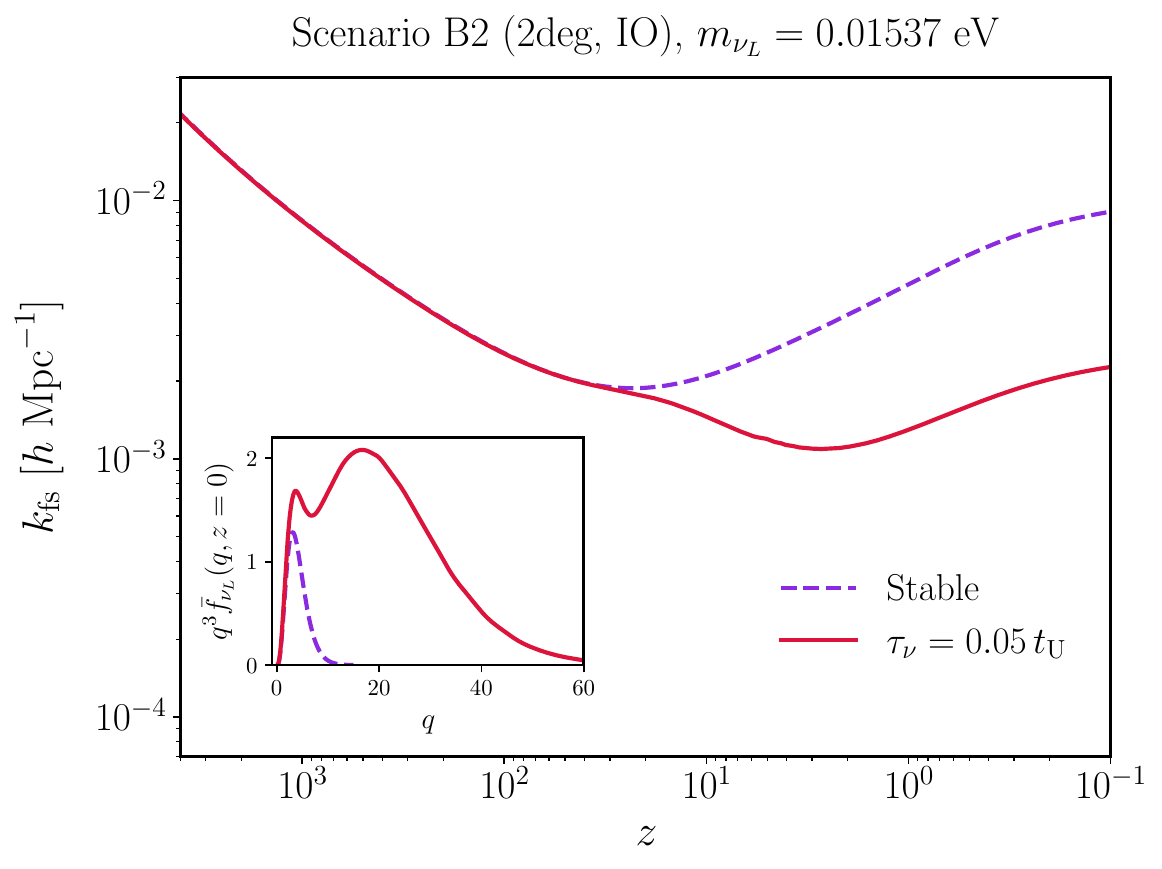}
\caption{Free-streaming scale of the lighter neutrino in scenarios B1 (\emph{left panel}) and B2 (\emph{right panel}), for neutrino decays with $\tau_\nu = 0.05 \,t_{\rm U}$ (solid lines) and the corresponding stable limit (dashed lines). In each case, the values of $\ml$ are chosen such that $\sum m_\nu =0.12\,\ev$. The inset plots display the final phase-space distribution of $\nu_L$ (solid lines), with comoving momentum $q$ given in units of $T_{\nu 0}$. These are compared to a standard Fermi-Dirac distribution (dashed lines).}
\label{fig:kFS_PSD}
\end{figure*}

The background Boltzmann equations to track the phase-space evolution of the $\{\nu_H,\nu_L,\phi\}$ system are given in Eqs. 4.12 - 4.14 of \cite{Barenboim:2020vrr}. In the non-relativistic limit (i.e., excluding inverse decay and quantum statistics terms), these equations reduce to 
\begin{align}
\frac{\partial \bar{f}_{\nu_H}\left(q_1\right)}{\partial \tau} &=-\frac{a^2 \mH \mathbb{g}_{H} \Gamma_\nu}{\epsilon_1} \bar{f}_{\nu_H}\left(q_1\right), \label{eq:BE_background_h} \\
\frac{\partial \bar{f}_{\nu_L}\left(q_2\right)}{\partial \tau} &=\frac{a^2 \mH^3 \mathbb{g}_{L} \Gamma_\nu}{\left(\mH^2-\ml^2\right) \epsilon_2 q_2} \int_{q_{1-}^{(\nu L)}}^{q_{1+}^{(\nu L)}} dq_1 \frac{q_1}{\epsilon_1} \bar{f}_{\nu_H} (q_1),
\label{eq:BE_background_l} \\ 
\frac{\partial \bar{f}_{\phi} \left(q_3\right)}{\partial \tau} &=\frac{2 a^2 \mH^3 \mathbb{g}_{\phi} \Gamma_\nu}{\left(\mH^2 - \ml^2\right) q_3^2} \int_{q_{1-}^{\phi}}^{\infty} dq_1 \frac{q_1}{\epsilon_1} \bar{f}_{\nu_H} \left(q_1\right),
\label{eq:BE_background_phi}
\end{align}
where $\epsilon_i \equiv \left(q_i^2+a^2 m_i^2\right)^{1 / 2}$, and the integration limits are given by 
\begin{align}
q_{1 \pm}^{(\nu L)} &=\left|\frac{\epsilon_2\left(\mH^2-\ml^2\right) \pm q_2\left(\mH^2+\ml^2\right)}{2 \ml^2}\right|,
\label{eq:q1_ml}  \\ 
q_{1-}^{(\phi)} &= \left|\frac{a^2\left(\mH^2-\ml^2\right)^2-4 \mH^2 q_3^2}{4 q_3\left(\mH^2-\ml^2\right)}\right|. 
\label{eq:q1_phi}
\end{align}
The factors $\mathbb{g}_{ i}$ account for the multiplicity of decay channels, and take the values $(\mathbb{g}_{H},\mathbb{g}_{L},\mathbb{g}_{\phi}) = (2,1,2)$ for scenario B1 and $(\mathbb{g}_{H},\mathbb{g}_{L},\mathbb{g}_{\phi}) = (1,2,2)$ for scenario B2.  In scenario A, both decay products $\{\nu_L,\phi\}$ are massless and can therefore be combined into a single DR fluid. In this way, \autoref{eq:BE_background_l}-\autoref{eq:BE_background_phi} can be  integrated over momenta, leading to the equation of motion for the background DR density \cite{Holm:2022eqq}, 
\begin{equation}
\frac{d\rho_{\rm dr}}{d\tau} +4aH\rho_{\rm dr} = a\Gamma_\nu \mH n_{\nu_H},
\end{equation}
where $n_{\nu_H}$ indicates the number density of $\nu_H$. In scenario B, $\nu_L$ is massive, so the momentum integration can only be performed for \autoref{eq:BE_background_phi}. This yields the modified DR equation \cite{Terzaghi:2025pvb}
\begin{equation}
\frac{d\rho_{\rm dr}}{d\tau} +4aH\rho_{\rm dr} = \varepsilon a\Gamma_\nu (\mathbb{g}_\phi/g_{H}) \mH n_{\nu_H},   
\end{equation}
where $\varepsilon \equiv \frac{1}{2} (1-\ml^2/\mH^ 2)$ accounts for the energy repartition between the decay products, and $g_H=1 (2)$ for B1 (B2). The corresponding first-order Boltzmann hierarchies for the neutrino and DR species are given in \cite{Terzaghi:2025pvb}.\ 

Using \autoref{eq:BE_background_h}-\autoref{eq:BE_background_l}, the PSD of $\nu_H$ and $\nu_L$ can be evolved from their initial Fermi-Dirac shape (as predicted by standard neutrino decoupling) until the present day. In the inset plots of \autoref{fig:kFS_PSD}, we show $\bar{f}_{\nu_L}(q,z=0)$ for cases B1 and B2, assuming $\tau_\nu = 0.05 \,t_{\rm U}$ and $\sum m_\nu =0.12\,\ev$. It can be seen that neutrino decays induce sizeable spectral distortions, over-populating the high-momentum tail of the lighter neutrino distribution. Using relativistic kinematics, the maximum comoving momentum at which $q^3\bar{f}_{\nu_L}(q,z=0)$ has support can be estimated as \cite{Barenboim:2020vrr}
\begin{equation}
q_{\rm max} = a_{\rm D} \varepsilon \mH + a_{\rm D} \frac{\mH^2+\ml^2}{\mH^2} v_{\nu H},
\label{eq:qmax_estimate}
\end{equation}
where $a_D$ is the approximate scale factor at which $\bar{f}_{\nu_H}$ has been fully depleted, and $v_{\nu H}$ is the velocity dispersion of $\nu_H$. Hence, the average momentum $\langle q\rangle$ is increased relative to a Fermi-Dirac distribution (for which $\langle q\rangle \sim 3 T_{\nu0}$), delaying the transition of $\nu_L$ to the non-relativistic regime. This effect is more prominent for scenario B2, since the $\nu_L$ distribution is sourced by two decaying $\nu_H$ particles. The increase in $\langle q\rangle$ directly impacts the evolution of the lighter neutrino free-streaming scale\footnote{The most important role is typically played by the largest free-streaming wavenumber, i.e. the one associated with the heaviest neutrino mass eigenstate. However, once $\nu_H$ has decayed away, it is the free-streaming scale of $\nu_L$ that determines the cutoff in $P_m(k)$.}
\begin{equation}
k_{\rm fs} = \sqrt{\frac{3}{2}} \frac{aH}{v_{\nu_L}}.    
\end{equation}
This scale is set by the velocity dispersion of $\nu_L$ \cite{Shoji:2010hm} 
\begin{equation}
v_{\nu_L}^2 (z) = \frac{\int dq q^2 (q^2/\epsilon_2^2) \bar{f}_{\nu_L}(q,z) }{\int dq q^2  \bar{f}_{\nu_L}(q,z)}.     
\end{equation}
On scales smaller than the free-streaming length, $k > k_{\rm fs}$, neutrino density perturbations undergo oscillatory behavior, leading to a suppression of matter growth. In the standard case, $k_{\rm fs}$ passes through a minimum $k_{\rm nr}$ at the time of the non-relativistic transition $z_{\rm nr} \sim \ml/\langle q \rangle $, after which it increases as $k_{\rm fs} \propto a^{1/2}$. As shown in \autoref{fig:kFS_PSD}, the late-time evolution of $k_{\rm fs}$ is modified by neutrino decays, extending the range of modes $k$ and redshifts $z$ over which matter perturbations are damped. \ 

The reduction in $k_{\rm nr}$ enhances the sensitivity of the CMB lensing potential---and consequently of the CMB power spectra---to neutrino masses, particularly for lifetimes in the region $\tau_\nu \sim t_{\rm U}$. As discussed in the main text, this leads to stronger constraints on $\sum m_\nu$ than suggested by naïve estimates, with virtually no relaxation in scenario B1 and even a tightening in scenario B2 relative to the stable case.\ 

\begin{figure*}[ht!]
\centering
\includegraphics[width=0.97\linewidth]{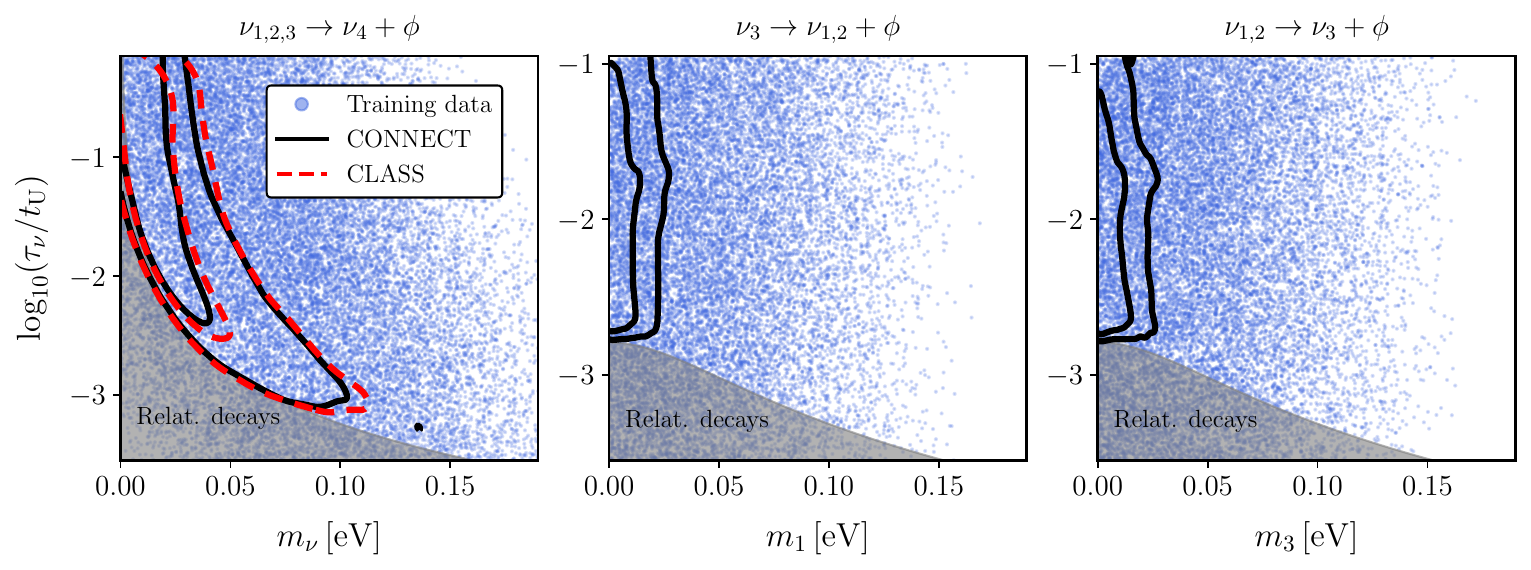}
\caption{Training data and \connect~posteriors in the $(\log_{10}(\tau_\nu/t_{\rm U}),\, m_{\nu})$--plane, for each of the three neutrino decay scenarios A (\emph{left panel}), B1 (\emph{middle panel}) and B2 (\emph{right panel}). For scenario A, we additionally show the posterior resulting from a standard \class-based run. The gray areas mark the regime of relativistic decays, which is excluded during the MCMC analysis.}
\label{fig:connect_training_data}
\end{figure*}

We note that previous studies have argued that cosmological neutrino mass bounds can be relaxed if neutrinos possess a larger average momentum than that of a thermal distribution \cite{Cuoco:2005qr,Oldengott:2019lke,Alvey:2021sji}. However, these works rely on parameterizations of the relic neutrino distribution that do not capture the time dependence or the shape of spectral distortions induced by neutrino decays of the type $\nu_i \rightarrow \nu_j + \phi$. As a consequence, such treatments lead to very different effects on $\rho_{\nu,0}$ and $k_{\rm fs}$, and hence to a different CMB phenomenology. Our results therefore indicate that an increased average neutrino momentum does not always imply a relaxation of cosmological neutrino mass bounds, with neutrino decays providing an important counterexample.

\begin{figure*}[ht!]
\centering
\includegraphics[scale=0.73]{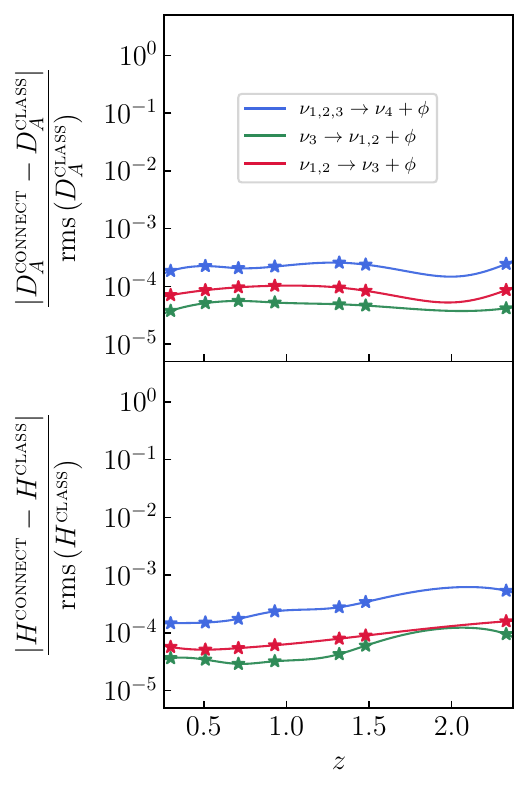}
\hspace{2mm}
\includegraphics[scale=0.73]{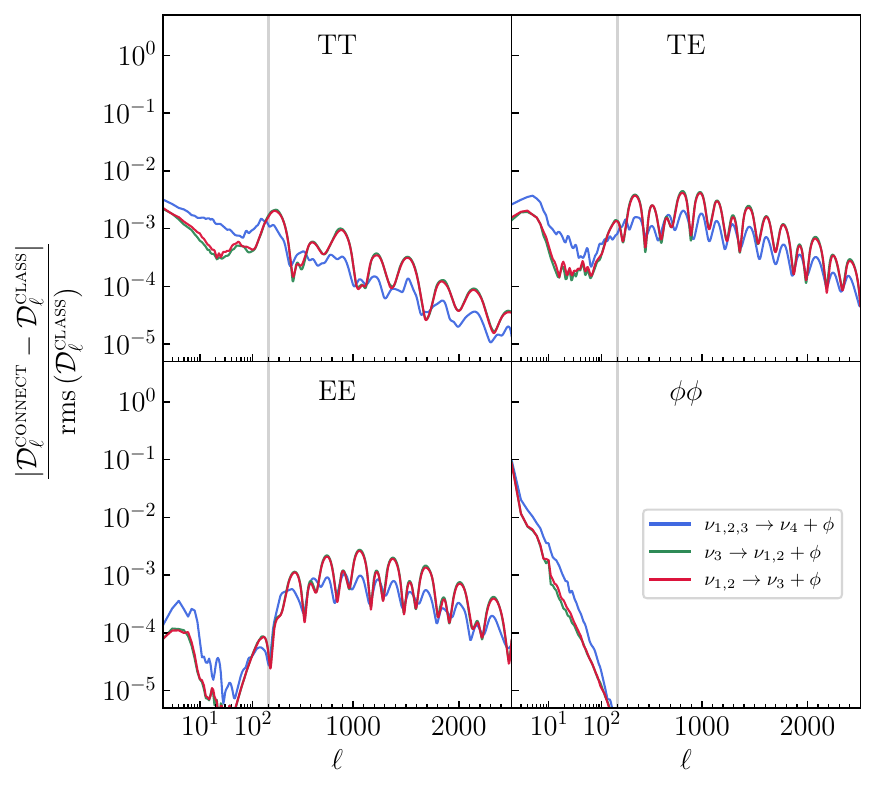}
\caption{Errors of the neutrino decay emulators in the background functions $(H(z),D_A(z))$ (\emph{left panel}) and CMB spectra (\emph{right panel}), evaluated on a test dataset of $\sim 10^3$ points. The stars mark the redshifts at which $H(z)$ and $D_A(z))$ have been emulated. All curves indicate the 95.45\% CL, meaning that 95.45\% of the test points have errors beneath the curves.}
\label{fig:connect_errors}
\end{figure*}

\section{Neural network emulators}\label{app:emulators}

We use \connect\ \,\cite{Nygaard:2022wri} to emulate the output of our augmented \class~version. By employing the hyperellipsoidal sampling scheme implemented in \connect, which concentrates training data in regions of higher likelihood \cite{Nygaard:2024lna}, we are able to obtain accurate emulators with  significantly fewer data points than traditional Latin hypercube approaches.\footnote{\connect’s standard iterative sampling approach is slightly more costly in terms of \class~evaluations; for our purposes, the hyperellipsoidal sampling scheme proved sufficient.}~The input layer of the neural network emulators is the parameter set $\{\omega_b, \omega_{\rm cdm}, H_0, n_s, \ln(10^{10}A_s), \tau_{\rm reio},m_\nu, \log_{10}(\tau_\nu/t_{\rm U})\}$, with $m_\nu =\mH$ for scenario A and $m_\nu  = \ml$ for scenarios B1/B2. The output layer comprises all  relevant cosmological observables: the CMB power spectra $C_\ell^{\rm XY}$ (with $\rm{XY=TT,TE,EE,} \,\phi\phi$) in the range $2 \leq \ell \leq 2500$, the Hubble rate $H(z)$ and the angular diameter distance $D_A(z)$ at the effective redshifts of DESI DR2 BAO, and various derived parameters, including the sound horizon at baryon drag, $r_s$. Regarding the precision settings, for the decay case A we use $N_q = 10$ momentum bins up to a maximum value $q_{\rm max}/T_{\nu 0} = 15 $, while for decay cases B1/B2 we implemented a routine (based on \autoref{eq:qmax_estimate}) that automatically determines the optimal values of $(N_q,\, q_{\rm max})$ for a given $\ml,\,\tau_\nu$ and mass ordering \cite{Terzaghi:2025pvb}. We truncate all Boltzmann hierarchies at $\ell_{\rm max} = 17$, and set $N_{\rm ur}:=0.00641$.

We use $2.5\times 10^4$ training samples for scenario A and $2\times 10^4$ samples for scenarios B1/B2. These samples were spread in 8-dimensional hyperellipsoids centered around the approximate best-fit points and with \lcdm~parameter correlations from previously converged MCMC runs.\footnote{For the model A, we also had information on the correlations for $(\log_{10}(\tau_\nu/t_{\rm U}),\, m_{\nu})$ from a previous \class-based MCMC.} The samples for the case A were generated in parallel in $\sim 1$ day using $128$ CPU cores, while the samples for the cases B1 and B2 required $\sim 3$ days on  $960$ CPU cores. The architecture of the neural networks is the same as in \cite{Nygaard:2022wri}. Each emulator was trained for $\sim 400-4000$ epochs with a batch-size of 512, which took $\sim 10-60$ min on two GPUs. \

In \autoref{fig:connect_training_data} we show the distribution of training samples together with the \connect\ posteriors for the three decaying neutrino models. The samples are visibly denser in regions where the posteriors have support. For scenario A, we also compare with a standard \class-based run ($\sim 8000$ CPU core-hours), finding excellent agreement with the \connect~result ($\sim 30$ CPU core-hours). We note that a \class-based MCMC for scenarios B1 and B2 is computationally unfeasible—in fact, this was the original motivation for employing emulators. In \autoref{fig:connect_errors} we show the $2\sigma$ percentile of the errors in the emulated observables. Following \cite{Nygaard:2022wri}, these errors are defined as the absolute difference between \class~and \connect~predictions scaled by the root-mean-square value of each observable. The errors generally remain well below $1\%$, providing strong validation of the emulator accuracy.\

\begin{figure}[h!]
\centering
\includegraphics[width=0.9\linewidth]{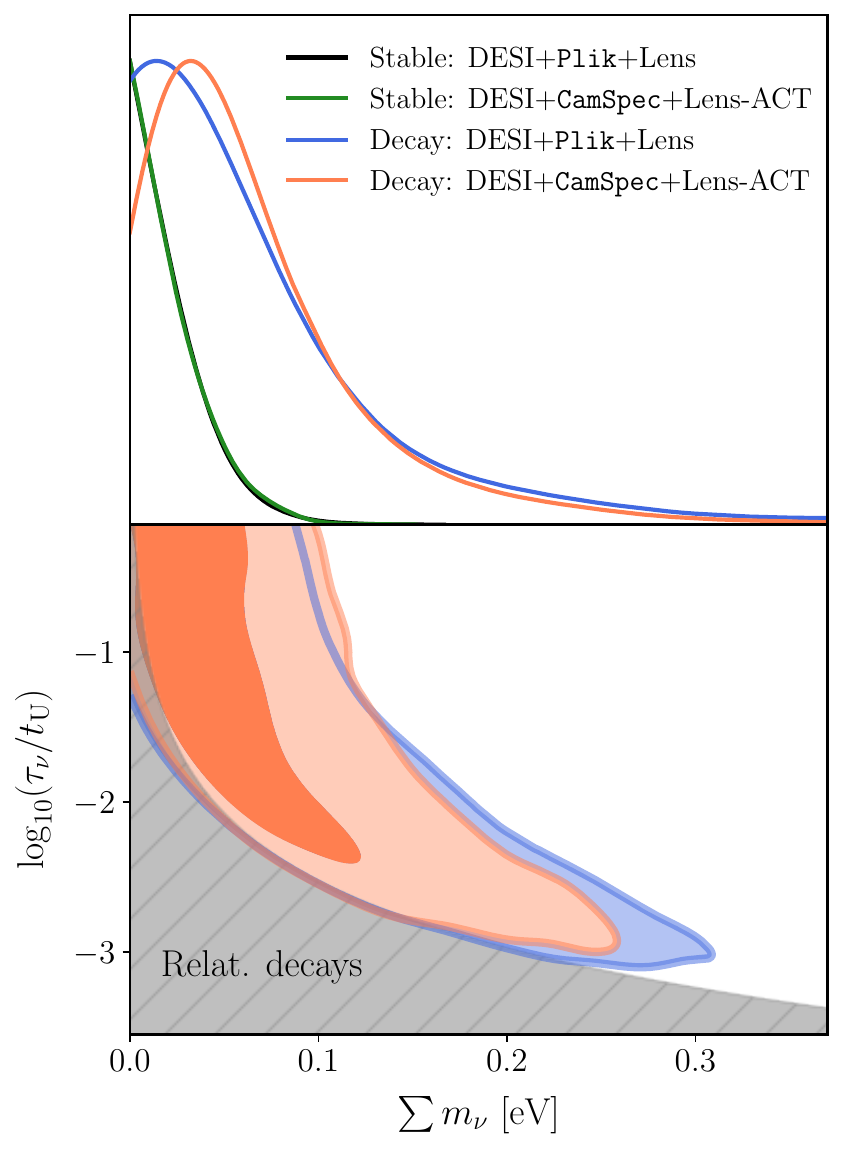}
\caption{Impact of variations in the CMB datasets on the total neutrino mass $\sum m_\nu$ and neutrino lifetime $\log_{10}(\tau_\nu/t_{\rm U})$, for neutrino decays into DR and the corresponding stable limit. We compare the \texttt{Plik} and \texttt{CamSpec} high-$\ell$ likelihoods, as well as the CMB lensing reconstruction from \Planck~PR3  and from the combined \Planck~PR4 + ACT DR6 data.
}
\label{fig:decay_DR_NPipe}
\end{figure}

\section{Impact of CMB likelihoods}\label{app:CMB_likelihoods}

To assess the robustness of our results, we investigated how our constraints are affected by the choice of CMB likelihoods. For concreteness, we focus on decays into DR and compute the theoretical predictions directly with \class, rather than using the \connect~emulators employed in the main text. We adopt an alternative set of CMB likelihoods chosen to match the baseline analysis by the DESI collaboration \cite{DESI:2025zgx}. In particular, we replace the \texttt{Plik} high-$\ell$ TTTEEE likelihood with \texttt{CamSpec}, which is built on the latest \texttt{NPIPE} \Planck~PR4 data release \cite{Rosenberg:2022sdy}. For the CMB lensing reconstruction, we replace \Planck~PR3 with the combined \Planck~PR4 + ACT DR6 likelihood from \cite{ACT:2023kun}. Given the precision of this CMB lensing likelihood, it becomes important to incorporate non-linear matter clustering. Hence, we use the non-linear code \texttt{HaloFit} \cite{Takahashi:2012em} as implemented in \texttt{CLASS}. \ 

The effect of the CMB likelihoods is summarized in \autoref{fig:decay_DR_NPipe}. Using \texttt{CamSpec} and \Planck~PR4 + ACT DR6 lensing, we obtain a neutrino mass bound for three stable degenerate neutrinos of $\sum m_\nu < 0.063 \, \ev$ (95\% CL), in good agreement with the baseline analysis by DESI DR2. When allowing decays into DR, the updated CMB likelihoods provide a tighter constraint than \Planck~PR3, yielding $\sum m_\nu < 0.19 \, \ev$ (95\% CL). While this represents a non-negligible shift, this bound could be slightly biased since \texttt{HaloFit} was not calibrated for decaying neutrino cosmologies. However, our main conclusion---that neutrino decays into massless BSM particles can reconcile cosmological and oscillation data---remains robust under variations in the CMB datasets.

\bibliography{references}

\end{document}